\documentclass{llncs}

\usepackage{amssymb,lscape}
\usepackage{graphicx}
\usepackage{enumitem}
\usepackage{multirow}
\usepackage{hhline}
\usepackage{url}
\usepackage{bbding} 
\usepackage{tabularx}
\usepackage{multirow}
\usepackage{booktabs}
\usepackage{xcolor}

\newcolumntype{L}{@{}l@{}} 

\urldef{\mailsa}\path|jinatamaradawood@gmail.com|
\urldef{\mailsb}\path|lucas.gren@getinge.com|
\urldef{\mailsc}\path||
\newcommand{\keywords}[1]{\par\addvspace\baselineskip
\noindent\keywordname\enspace\ignorespaces#1}

\begin{document}

\mainmatter  
\title{Prototypical Leadership in \\Agile Software Development}

\titlerunning{Prototypical Leadership in Agile}

\author{Jina Dawood\inst{1} \and Lucas Gren\inst{1,2}
}
\authorrunning{Jina Dawood and Lucas Gren}

\institute{Chalmers University of Technology $|$ University of Gothenburg, \\
Gothenburg, Sweden\\ 
\and
Getinge AB\\
Gothenburg, Sweden\\ 
\mailsa\\
\mailsb\\
\mailsc\\}

\toctitle{Lecture Notes in Computer Science}
\tocauthor{Authors' Instructions}
\maketitle

\begin{abstract}
Leadership in agile teams is a collective responsibility where team members share leadership work based on expertise and skills. However, the understanding of leadership in this context is limited.  This study explores the under-researched area of prototypical leadership, aiming to understand if and how leaders who are perceived as more representative of the team are more effective leaders.  Qualitative interviews were conducted with eleven members of six agile software teams in five Swedish companies from various industries and sizes.  In this study, the effectiveness of leadership was perceived as higher when it emerged from within the team or when leaders aligned with the group. In addition,  leaders in managerial roles that align with the team's shared values and traits were perceived as more effective, contributing to overall team success.  

\end{abstract}

\section{Introduction}\label{ref:intro}
Leadership and management, while often conflated, are distinct concepts in organizational theory. ``Leadership [involves] the articulation of an organizational vision, introduction of major organizational change, providing inspiration, and dealing with highly stressful and troublesome aspects of the external environments of organizations''~\cite[p. 444]{house1997social}. This definition emphasizes the visionary and inspirational aspects of leadership.
In contrast, ``Management consists of implementing the vision and strategy provided by leaders, coordinating and staffing the components of organizations, administering the infrastructures of organizations, and handling the day-to-day problems that inevitably emerge in the process of strategy and policy implementation and ongoing organizational functioning''~\cite[p. 445]{house1997social}. Management, therefore, focuses on the operational aspects of realizing the leader's vision.
Distilling these concepts further, ``leadership is getting group members to achieve the group's goals''~\cite[p. 315]{hogg2014sp}. This succinct definition captures the essence of leadership: influencing and guiding others toward a common objective. \textit{Leadership is a kind of work.} It is important to note that even in self-organized and self-managed agile teams, leadership is still present \cite{agileleadicse}. The distinction lies in how it is distributed; rather than being confined to a single individual, many team members can dynamically share the leadership work, collaboratively driving the team toward its goals \cite{agileleadicse,moe2009understanding}. Team members and managers share leadership work based on their expertise and skills, ensuring equal participation and decision-making  \cite{langfred2000paradox,agileleadicse}. 		

Agile software development emphasizes collaboration, flexibility, iterative development and customer satisfaction \cite{fowler2001}. By embracing agile practices and methods, organizations can increase productivity and deliver higher-quality software products to their customers \cite{moe2009understanding}.

Self-managing teams face challenges like conflicting priorities and opinions, which impact their performance \cite{tessem2014individual}. Effective leadership remains essential in guiding and supporting agile teams to achieve their goals, whether shared or formal \cite{spiegler2021empirical}. However, the transition from traditional to dynamic leadership has created confusion regarding the role of leadership in agile teams \cite{xu2018role}. In addition, limited research on leadership in agile teams makes it hard to grasp what makes a leader effective, even though leadership is vital for successful self-organization \cite{spiegler2021empirical}.				

Existing studies on leadership in agile software development have predominantly concentrated on leadership styles and practices, overlooking the broader and more complex aspects of leadership \cite{agileleadicse,moe2009understanding,modi2020leadership}. A systematic literature review on agile leadership stated that ``agile leadership research needs further attention and that more empirical studies are needed to understand agile leadership in general better.'' \cite{modi2020leadership}. A recent study by Gren et al.~\cite{agileleadicse} highlights the necessity for additional research on leadership in agile software development. In addition, the authors propose that exploring prototypical leadership could be valuable.

Prototypical leadership refers to leaders who exemplify their group’s highly valued characteristics and values \cite{kark2013dual}.   To the best of our knowledge, no existing research explicitly examines prototypical leadership in the context of agile software development. However, research in other fields suggests that prototypical leadership can be highly beneficial. For example,  Avolio et al.~\cite{avolio2022leader} found that prototypical leadership attributes are essential in influencing transformational leadership ratings. Leaders who exhibited prototypical characteristics were rated higher than those who showed anti-typical features.  Grille et al.~\cite{grille2015promoting} also found that promoting prototypical team leader behavior can increase psychological empowerment and shared leadership within teams.

This study aims to explore prototypical leadership through the perspectives of individuals in the industry. Ultimately, this study aims to gain an initial understanding of prototypical leadership in agile software development. The research questions addressed in this study are:

RQ1: How do members of agile software development teams perceive and experience leadership that aligns with team characteristics and values (i.e.\ prototypical leadership)?

RQ2: What impact does prototypical leadership have on agile teams?

Here, agile software development teams refer to the ideology in the Agile Manifesto \cite{fowler2001} and the practices and methods that reflect those principles, such as scrum \cite{diebold2014agile}. We have decided to center our research on the experiences and perspectives of individuals working within the industry. Therefore, we  collected qualitative data through semi-structured interviews to answer the research questions. Eleven members of six agile teams in five organizations in Sweden were interviewed. The qualitative data was analyzed using thematic analysis.

This paper is organized as follows: Section~\ref{ref:rw} examines theories and existing research related to our objectives. Then, in Section~\ref{ref:method}, we describe our research design. In Section~\ref{ref:results} we report our findings. Section~\ref{ref:discussion} addresses our research questions, discusses the findings with existing literature, and explores their implications. Finally, we conclude the paper and provide some views on future research in Section~\ref{ref:conclusion}.

\section{Related Work}\label{ref:rw}
This section provides an overview of the theoretical foundation for our study on prototypical leadership in agile software development teams. We focus on the Social Identity Theory of Leadership \cite{hogg2001social} and its application to agile contexts.

Social Identity Theory, introduced by Tajfel et al.~\cite{tajfel1971social}, explains how individuals categorize themselves and others into social groups based on shared traits, values, and norms. This categorization influences an individual's behavior and self-concept\footnote{Self-concept refers to an individual's perception of themselves, including their attributes, abilities, and identity within social groups \cite{hogg2001social}.}. Building on this foundation, Hogg et al.~\cite{hogg2001social} developed the Social Identity Theory of Leadership, which emphasizes the leader's role as a group member and their ability to embody and promote the group's identity. According to this theory, effective leaders are those who best represent the group's defining characteristics and values, known as prototypical leaders.

Prototypical leadership refers to leaders who exemplify their group's highly valued characteristics and values \cite{kark2013dual}. These leaders are seen as embodying the group's identity and are often more effective in gaining support and influencing team members \cite{van2003social}.
Research has shown that prototypical leaders often receive more support from their team members and are perceived as more effective \cite{barreto2017evaluation,nye1991effects,rast2012leadership}. This is based on the belief that a prototypical leader will act in the team's best interest and contribute to the group's overall success.

In agile software development, leadership often takes on a more collaborative and distributed form. Gren et al.~\cite{agileleadicse} found that leadership in agile teams is dynamically shared among team members, fostering a sense of team belonging and adapting to different organizational cultures.
While research on prototypical leadership in agile contexts is limited, studies on shared leadership in agile teams suggest potential connections. Moe et al.~\cite{moe2009understanding} highlighted that shared leadership positively impacts team performance and member satisfaction by enabling collaboration and effective decision-making.
Our study aims to bridge the gap between prototypical leadership theory and the practical realities of leadership in agile software development teams.

\section{Method}\label{ref:method}
This study aims to explore the concept of prototypical leadership in agile software development trough a qualitative interview study. Rather than focusing on quantitative measurements or quantities, we wanted to understand the ``how'' and ``why'' aspects \cite{runeson2009guidelines}. 
While we collected data from multiple teams, our primary unit of analysis is the individual participant's perspective rather than the team or organization as a whole. Therefore, this study is best characterized as an interview study rather than a multiple case study.

Qualitative data was collected through one-to-one interviews with eleven participants from six agile software teams in five Swedish companies.  Thematic analysis \cite{braun2006using} was employed to analyze the collected qualitative data. 
The following section will elaborate on our research design and the rationale behind our choices.

\paragraph{Cases}\label{ref:cases}
The cases in this study involved five Swedish companies selected based on their agile software development teams. These companies represented different industries and varied in size, offering various perspectives. Some of these companies operated internationally, while the teams being studied were all located in Gothenburg, Sweden. One of the companies included in the study was a multinational corporation operating in the automotive sector, from which two teams were studied. Another company operated in the food and beverage industry. The study also included a digital solutions provider, a small startup with nine employees, and a company specializing in disruptive technologies in the automotive sector. To maintain anonymity, no further details can be disclosed about these organizations.

\paragraph{Participants}\label{ref:participants}
We employed a combination of convenience and purposive sampling strategies to select participants \cite{etikan2016comparison}. Our sampling approach aimed to ensure diversity in our dataset while meeting specific criteria: (1) Participants working within agile software development teams, and (2)Representation from organizations of different sizes and industries. Additionally, we sought to include participants with varying ages, genders, and professional backgrounds to capture a range of perspectives.


Nine men and two women between 24 and 61 years old participated. Six participants were selected from the same team, as all individuals expressed a will to participate. Additionally, two were from different groups within the same company, while the remaining three were from different companies. All participants were part of software development teams, although not all held development roles. Some participants had transitioned into management roles but had previous experience as developers.    				

Data saturation \cite{guest2006many} was reached after the ninth interview. However, we conducted the two extra initially planned interviews to verify the data saturation. To ensure the participants' anonymity, no additional details beyond what is presented will be disclosed, as individuals may be identified (see Table~\ref{teaminfo}).

\begin{table}  

  \centering
  \caption{Team Information}
  \begin{tabular}{cccc}
    \toprule
    ID & Team & Team Size & Industry \\
    \midrule
    A1 & A & 6 & Automotive \\
    B1 & B & 5 & Automotive \\
    C1 & C & 8 & Software Consultancy \\
    C2 & C & 8 & Software Consultancy \\
    C3 & C & 8 & Software Consultancy \\
    C4 & C & 8 & Software Consultancy \\
    C5 & C & 8 & Software Consultancy \\
    C6 & C & 8 & Software Consultancy \\
    D1 & D & 13 & Retail \\
    E1 & E & 8 & Automotive \\
    F1 & F & 7 & Foodservice \\
    \bottomrule
  \end{tabular}
\label{teaminfo}
\end{table}

\paragraph{Data Collection}\label{ref:datacollection}
We gathered qualitative data through one-to-one interviews, which enabled us to obtain an in-depth understanding of the constructs. The interviews were semi-structured with mostly open-ended questions, allowing for a flexible conversation that helped us comprehensively understand the participants' reasoning \cite{braun2006using}.

An initial interview guide was developed based on a literature review on the topic and the stated research questions.					

We first conducted two pilot interviews with industry professionals to identify potential issues and areas for improvement in the interview guide. During the first pilot interview, we realized that the interview guide contained too many close-ended questions, limiting the conversation's flexibility. Therefore, we modified our guide and conducted a second pilot interview to ensure our changes addressed the issue. After the second round, we realized some inconsistencies in the wording of specific questions, which we modified. The interview guide is available at Zenodo\footnote{ \url{https://zenodo.org/records/13769850}}. 				

The pilot interviews were conducted solely in Swedish, as eight out of eleven participants were scheduled for interviews in Swedish. However, the interview guide was translated carefully to ensure consistency across all interviews. 						

The modified interview guide used to collect qualitative data contained seventeen mainly open-ended questions. The interview guide comprised four sections: an introduction that included questions to gather demographic information, questions grouped into two themes (Team and Leadership), and a closing section where participants could provide additional comments. The data collected during the pilot interviews were not included in the data analysis and were not considered part of the  final sample of eleven participants.

Face-to-face interviews were conducted at the participants' workplaces, except for one individual who had a remote interview due to logistical constraints. Before the interviews, we stated that the study would focus on team dynamics and leadership. The participants were also provided with information regarding the use of their data, the voluntary nature of their participation, and the confidentiality of their responses. Finally, with the participant's consent, the interviews were audio recorded to ensure the accuracy of the data collected. The interviews lasted between 45 and 80 minutes and took place over three weeks. 			

Eight interviews were held in Swedish, while three were conducted in English. Interview materials were translated carefully to maintain consistency. The transcripts obtained from these interviews form the primary source of qualitative data. However, we also took notes during the interviews to highlight important parts and assist in guiding the conversation more effectively. 

Descript\footnote{ \url{https://www.descript.com/}} was used to transcribe the interviews, and then we manually evaluated them to ensure accuracy. In addition, extraneous filler words were removed to make the transcripts more readable. Finally, all transcripts originally in Swedish were translated into English to simplify the coding process. We aimed at making the translation as accurate as possible by using DeepL\footnote{ \url{https://www.deepl.com/translator}}  and manually checking the accuracy. We provided the participants with a summary of their statements for verification, which they confirmed. 
One researcher transcribed all interviews. To keep everything organized and confidential, we gave each transcript and participant a unique number (see Table 1). We have taken an inductive approach in the data analysis, allowing the data to guide and shape our conclusions rather than beginning with preconceived notions \cite{braun2006using}. 
The data were analyzed using the 6-step thematic analysis method proposed by Braun and Clarke \cite{braun2006using} in ATLAS.ti\footnote{ \url{https://atlasti.com/}}. Since our research focuses on understanding the experiences and perspectives of our participants, thematic analysis is particularly appropriate for our study. The process included the following steps: (1) Familiarizing with the data: the transcripts were reviewed multiple times. (2) Generating initial codes: interesting phrases or segments were highlighted by assigning codes. (3) Searching for themes: codes that appeared frequently and could be categorized into broader themes were grouped.  (4) Reviewing themes: themes were reviewed, and a decision was made on whether to keep or drop them. (5) Defining and naming themes: names were given to the themes that reflected their content. (6) Producing the report: in section 4 of the report, the main findings and quotes supporting each theme are presented.

      The analysis followed an iterative process; steps 2-5 were repeated multiple times until thematic saturation was achieved \cite{braun2006using}. Thematic saturation was considered achieved when no new themes or meaningful insights emerged from the data. 

\begin{table*}
  \centering
  \caption{Example Theme, Codes, and Quotation}
  \begin{tabular}{p{3cm}p{4cm}p{5cm}}
    \toprule
    Theme & Codes & Quotation \\
    \midrule

    Leader-Team Alignment for Success & Group identity and values Representation, Leadership reflection & \textit{``Because as a leader of a group you should be a reflection of the group in some way. Yes, it is like a church tower. You have the whole church, but it is like the church tower that stands out and the one everyone sees first and it kind of points to everything, the whole foundation.''} [C3] \\
    \bottomrule
  \end{tabular}
\end{table*}

\section{Results}\label{ref:results}
We identified three themes that emerged from our analysis: (1)  \textit{Shared Leadership Based on Tasks and Expertise, (2) } \textit{Informal Hierarchy Based on Seniority,} and (3) \textit{Leader-Team Alignment for Success.}

\paragraph{Shared leadership based on tasks and expertise}
    In agile teams, participants emphasized that leadership is shared among team members, with different individuals taking on informal leadership roles based on their tasks and expertise. This approach allows each member to take responsibility and contribute as a leader in their expertise, which enables many team members to conduct leadership. If many team members share the leadership work, the leadership is perceived as stemming from the team members and thus being more prototypical. As one participant expressed:

 \begin{quote}
\textit{``So, everyone has sort of an informal leadership in the task really all the time and then you have to try to adapt it to the person and what they need. ''} [C4]
 \end{quote}

     The participants’ emphasized that the team's success rarely depends on just one person. Instead, everyone in the team works together and contributes to achieving common goals. Decisions are made by considering everyone's input and expertise. As one participant said:

\begin{quote}
\textit{``But when it comes to implementation, those things, then I feel like it’s more the team.Because everyone kind of brings different expertise to the table. ''} [C5]
 \end{quote}

     In agile teams, leadership is a collective effort encompassing everyday interactions and knowledge sharing among all team members. It involves everyone contributing their expertise and supporting one another, regardless of their roles or tasks. Multiple participants expressed this understanding of leadership:

\begin{quote}
\textit{``That we stand up and help each other and teach each other. That is what makes us reach our goal.''} [B1]
 \end{quote}

\paragraph{Informal hierarchy based on seniority}
  In agile teams, senior members or those with extensive experience often hold significant influence, as most participants recognized. They have a deep understanding of specific technologies and development approaches, contributing to successful project implementation. These senior members are defined by their prototypicality and often represent the values and norms of the team. One participant highlighted the impact of senior leaders, stating:

\begin{quote}
\textit{``Some of the best developed things are based on one of the most senior people taking the lead on something. And really got people on board with why you should have a certain technology or why you should develop this and that.''} [F1]
 \end{quote} 

Participants emphasized that senior members not only offer technical expertise but also actively share their extensive experience and guidance. As one participant highlighted:

\begin{quote}
\textit{``[name] has been developing for [many] years so [name] is also here to spread their knowledge to everyone else.''} [C5]
 \end{quote}

     Moreover, participants recognized that the perception of seniority in the workplace often means that their opinions are seen as more important and receive greater attention.  As one participant explained:

\begin{quote}
\textit{``Seniority is often such a thing in the workplace that you listen to the person who has the experience or the person who has been doing the project the longest. That their opinions are somehow more valuable or and that they are heard more anyway.''} [E1]
 \end{quote}

\paragraph{Leader-team alignment for success}
Most participants preferred team members as managerial leaders, despite potential concerns about group friction. They value leaders who know the team and their work, as highlighted by one participant:

\begin{quote}
\textit{``If you are in a role that can affect the job or my environment positively or negatively. I would rather have someone who knows what they do and what they work with. And who knows me or something like that and you really do if we have worked together.''} [E1]
 \end{quote} 

Regardless of whether the leader comes from within the team or externally, all participants stressed the significance of the leader aligning with the team's values and working methods without imposing unnecessary changes. They stressed that a leader should represent and embody the values and essence of the group they lead. One participant [C3] compared a leader to a church tower that symbolizes and stands out for the entire church, highlighting the significance of this representation.	          

      Participants also stressed the significance of leaders actively engaging with the team, including sharing meals and informal interactions. They prefer leaders who are present and involved, unlike those who lack meaningful contact with the team. One participant shared their preference:

   \begin{quote}
\textit{``We usually eat lunch together. And I feel like that is important. I like that a lot better than someone coming in telling me what to do and then leaving and there's no real interaction.''} [C5]
 \end{quote}

     The findings show some possible positive outcomes such as job satisfaction and reduced turnover when leaders have aligned with their team. One participant shared their experience about an external leader who came into the team:

   \begin{quote}
\textit{``So, if we have a natural turnover of staff and then you suddenly get a leader who is liked and everything works well and no one is leaving because they like it, that's an indicator. The only people who will leave are those who want to move on to some other development of some kind.''} [A1]
 \end{quote}

     On the other hand, leaders who are perceived as unproductive often struggle to align themselves with the team. This lack of alignment erodes trust, hindering effective teamwork. One participant shared their experience:

   \begin{quote}
\textit{``Since he wasn't in the group at all, no one trusted him, and he doesn't trust the group either because he wasn't in the group. So, he was constantly checking that everyone is doing the right thing.''} [C2]
 \end{quote} 

The findings highlight the nuanced nature of leadership in agile teams. Success in leadership extends beyond technical skills, requiring alignment with the team and active engagement. This alignment seems to contribute to positive collaboration, increased job satisfaction, and potentially lower the turnover.

\paragraph{Summary}\label{ref:summary}

 \begin{enumerate}
\item Since agile teams share leadership work, the leadership is perceived as more prototypical.
\item Not only one leader but the senior members of agile teams take on more leadership work, meaning it is also more prototypical then if only having one senior person who leads. 
\item Effective managerial leadership in agile teams goes beyond technical expertise and requires alignment with the team's goals and values, also showing the positive effects of prototypical leadership.
 \end{enumerate}

\section{Discussion}\label{ref:discussion}
\paragraph{RQ1 How do members of agile software development teams perceive and experience leadership that aligns with team characteristics and values (i.e.\ prototypical leadership)?}\label{ref:r1}
Our research reveals the perception of prototypical leadership in agile software development in various ways. To comprehend these findings and address RQ1, we draw upon two theories: Social Identity Theory \cite{tajfel1971social} and Social Identity Theory of Leadership \cite{hogg2001social}. The former theory explains the formation of agile teams based on shared traits, norms, and values, mirroring the categorization of individuals into groups. The study aligns with this theory, emphasizing that team formation centers on shared traits, norms, and values.

As team members collectively define their identity in relation to the group, they also become more representative of the group, a concept known as group prototypicality \cite{hogg2001social}. This aligns with the Social Identity Theory of Leadership, where group members depend on leaders who embody the group's traits and values, attracting followers and support.

The above results show that, in agile teams, many leaders are likely to embody leader prototypicality since the leadership is more shared. Team members who assume leadership positions tend to exhibit the defining traits of the group, becoming prototypical leaders  \cite{kark2013dual}. This phenomenon can also extend to managerial roles within the team, as senior members often progress to managerial positions. Additionally, external managerial leaders who share values and traits with the team tend to exhibit prototypical leadership.  This concept may explain why teams accept some appointed leaders but reject others. A potential conflict can arise between various types of leaders within a team: those who emerge naturally and embody the team's ideal characteristics (prototypical), and formally appointed managers. To mitigate this conflict, it is advisable to ensure that an appointed manager's values and traits align closely with those of the team. However, if the team's emerging values and norms diverge significantly from the organization's, it becomes crucial to proactively establish team norms during the team formation process. This approach can help bridge the gap between organizational goals and team dynamics, fostering a more cohesive and effective work environment.

\paragraph{RQ2 What impact does prototypical leadership have on agile teams?}\label{ref:r2}	
This study highlights the significant role of prototypical leadership in agile software development teams, bridging theoretical foundations with practical realities. Our findings reveal that team members who embody the characteristics of prototypical leaders facilitate shared leadership by aligning with the team's values, norms, and shared traits. Drawing from the Social Identity Theory of Leadership \cite{hogg2001social}, we observe that these leaders play a crucial role in shaping and reinforcing the team's social identity. In the context of agile development, where leadership is often distributed and collaborative \cite{agileleadicse}, prototypical leaders emerge as key figures who embody the team's identity, contributing to success by sharing knowledge, guiding peers, and supporting team members. This aligns with Moe et al.'s \cite{moe2009understanding} findings on the positive impact of shared leadership on team performance and member satisfaction in agile environments.

Our study reveals that senior members within agile teams often assume the role of prototypical leaders, retaining significant influence and naturally attracting followers. This phenomenon can be explained through the Social Identity Theory of Leadership, which posits that effective leaders are those who best represent the group's defining characteristics \cite{hogg2001social}. The support these senior members receive aligns with previous research indicating that prototypical leaders often receive backing even when their actions might be perceived as unfair \cite{ullrich2009substitutes,van2003social}. This unconditional support stems from the belief that prototypical leaders act in the team's best interest \cite{kark2013dual}, a perception reinforced by their track record of contributing to the team's success. In our observed agile team context, senior members significantly influence both technical work and social identity formation.

The presence of prototypical leaders in agile teams appears to have a positive impact on team dynamics and performance. This impact can be attributed to enhanced group-oriented behavior, which van Knippenberg and Hogg \cite{van2003social} identify as crucial for leadership effectiveness in organizations. In agile teams, this translates to improved collaboration and a stronger focus on collective goals. Furthermore, the alignment of prototypical leaders with team values and norms facilitates more effective decision-making processes, supporting Moe et al.'s \cite{moe2009understanding} findings on the positive impact of shared leadership on team performance. Our study suggests that the presence of prototypical leaders contributes to higher job satisfaction among team members, which can be attributed to the leader's ability to represent and promote the group's identity, as highlighted by the Social Identity Theory of Leadership \cite{hogg2001social}.

In conclusion, prototypical leadership seems to have a positive impact on agile teams. Agile teams benefit from leaders who align with the team's values and exhibit group-oriented behavior, which then seems to lead to higher job satisfaction.


\subsection{Implications}\label{ref:impl}
The study's findings carry some implications for bolstering leadership effectiveness within agile software teams. Organizations aiming to cultivate teamwork and effective leadership could prioritize individuals who align well with the team. While it is crucial for organizations to seek individuals who harmonize with agile teams, it is equally important to remain mindful of the potential limitations of exclusively focusing on this aspect. Striking a balance between aligning team members and recognizing the value of diverse skills and backgrounds could yield benefits.

Leaders, particularly senior members, should exemplify team-focused behavior and possess the requisite attributes for managerial roles. Actively engaging with the team, making collective decisions, and adhering to agile software development principles can enhance senior members' support within the team. Finally, organizations could strive to encourage prototypicality and group-oriented behavior among their leaders to augment effective leadership. This entails selecting leaders who share alignment with the team's values, norms, and traits, while also supporting their development in embodying the group's defining characteristics. By taking these implications into account, organizations might be able to foster more effective leadership.

\subsection{Threats to validity}\label{ref:vts}
This section addresses potential validity threats in the study, encompassing four key areas: external validity, internal validity, construct validity, and reliability, following parts of the framework established by Gren \cite{gren2018standards}. 

\paragraph{External validity}
The study's limited sample size, drawn from teams exclusively in Sweden, raises concerns about the generalizability of the findings. The use of convenience and purposive sampling methods introduces the potential for bias and restricts the broad applicability of the results. Moreover, the inclusion of six participants from the same team may introduce bias and might not accurately represent diverse teams in different organizations.

To somewhat mitigate these threats, we made deliberate efforts to enhance sample diversity by selecting participants with varying backgrounds, teams, ages, genders, and roles. Additionally, the study aimed to encompass organizations from diverse industries and sizes to achieve a more representative sample.

It is also important to note that our study was conducted exclusively in Swedish companies, which may limit the generalizability of our findings. National, regional, and cultural contexts can strongly influence values and norms related to leadership and management styles. Sweden, in particular, is known for its preference for flat work hierarchies and participative leadership \cite{holmberg2006modelling}.

\paragraph{Internal validity}
The reliance on self-reported data in the study may be influenced by recall bias and social desirability bias. To try to address these biases, we emphasized honesty, anonymity, and voluntary participation. Conducting interviews in both Swedish and English could also have impacted the quality of conversations, introducing potential language-related biases and variations in participant comfort levels and fluency.

\paragraph{Construct validity}
A qualitative interview study and a thematic analysis method inherently involve researcher interpretation and judgment, introducing subjectivity. As a single researcher conducted the interviews and coding,  personal biases and preconceived notions may have influenced data collection, analysis, and result interpretation. To try to address this threat, the second author reviewed the quotes and their related codes. 

\paragraph{Reliability}
The translation process from Swedish to English for the interview guide and transcripts introduces the possibility of errors or inaccuracies, potentially affecting the reliability of data analysis. To somewhat mitigate this, we used software for initial translation and conducted manual checks. Additionally, participants received summaries of their statements to confirm accurate interpretation.

\section{Conclusion and Future Work}\label{ref:conclusion}
Our study explored prototypical leadership in agile software development teams through qualitative interviews with six teams across five Swedish companies. We found that prototypical leadership does exist in agile software development, and possibly even more than in a tradition team setup due to shared leadership and senior members also doing leadership work.  We also saw that protptypicality is seen as increasing the leaders effectiveness. 

Future research could delve into the dynamics and implications of prototypical leadership in agile teams. Investigating preferences for internal leaders and their impact on group cohesion and prototypicality, as well as exploring how follower perceptions influence leadership effectiveness, are valuable areas for further study. Also, future research should explore how prototypical leadership manifests in agile teams across different cultural contexts.

\bibliographystyle{splncs}
\bibliography{ref}
\newpage

\end{document}